\documentclass[12pt]{article}
\usepackage{graphicx}
\usepackage{dcolumn}
\usepackage{bm}
\usepackage{amssymb}
\usepackage{amsmath}
\newcommand{\bq}{\begin{equation}}
\newcommand{\eq}{\end{equation}}

\newcommand{\bqr}{\begin{eqnarray}}
\newcommand{\eqr}{\end{eqnarray}}

\newcommand{\bqrx}{\begin{eqnarray*}}
\newcommand{\eqrx}{\end{eqnarray*}}

\newcommand{\br}{\begin{array}}
\newcommand{\er}{\end{array}}

\newcommand{\voo}{\vspace*{5pt}}

\begin{document}

\pagestyle{empty}

\setlength{\parindent}{18pt}
\setlength{\footskip}{.5in}



\vspace*{.6in}
\begin{center}
Asymptotic distributions of quantum walks on the line with two entangled coins
\end{center}
\begin{center}
Chaobin\, Liu\\
Department of Mathematics\\
Bowie State University\\
 Bowie, MD 20715 USA\\ 
{\footnotesize \,\,cliu@bowiestate.edu}
\voo\\
\voo\voo\voo
\end{center}
\begin{abstract}
We advance the previous studies of quantum walks on the line with two coins. Such four-state quantum walks driven by a three-direction shift operator may have nonzero stationary distributions (localization), thus distinguishing themselves from the quantum walks on the line in the basic scenario (i.e., driven by a single coin). In this work, asymptotic position distributions of the quantum walks are examined.   We derive a weak limit for the quantum walks and explicit formulas of stationary probability distribution, whose dependencies on the coin parameter and the initial state of quantum walks are presented. In particular, it is shown that the weak limit for the present quantum walks can be of the form in the basic scenario of quantum walks on the line, for certain initial states of the walk and certain values of the coin parameter. In the case where localization occurs, we show that the stationary probability exponentially decays with the absolute value of a walker's position, independent of the parity of time. 

\end{abstract}

keywords: quantum walks with two coins, three-direction shift operator, stationary distribution, localization, weak limit.

\section{Introduction}	

Quantum walks (QW) may be described as the natural counterparts of classical random walks, governed by the principles of quantum mechanics.
Like classical random walks, QW are classified into two main types: discrete-time QW \cite{NV00,ABNVW01,AAKV01} and continuous-time QW \cite{FG98,CFG02}. In this paper, we study the discrete-time case. Let $\mathbb{Z}$ be the set of all integers, the QW on $\mathbb{Z}$, has been extensively explored. Among all the findings are the weak limit theorems, such as \cite{K02,K2005,GJS04}.

Venegas-Andraca et al. \cite{V-ABBB05} introduced and investigated the QW on $\mathbb{Z}$ with two entangled coins. For this 4-state QW, its conditional shift operator embraces three distinct directions. At every time step of the walk, depending on the state of the coin, the walker moves forward, or moves backward, or stalls at that step. 
In their numerical simulations of the QW \cite{V-ABBB05}, the authors found the phenomenon, called ``localization", whereby the probability distribution of the walker's position is seen to exhibit a persistent major ``spike" (or ``peak" ) at the initial position. In what followed,  Liu and Petulante \cite{LP09} gave theoretical explanations for the observation of ``localization". By the analysis of the spectral properties of the evolution operator for the QW, a general formula was derived for the limiting probability, from which the authors deduced the limiting value of the height of the observed spike at the origin. In this work, we advance the study of this model of QW driven by a parameterized coin instead of the Hadamard coin. Using the method of the Fourier transform, we show that the normalized operator $\frac{1}{t}X_t$ converges weakly to a random variable as $t\rightarrow \infty$. The limit measure is described by both a $\delta$-function corresponding to localization and a density function. As our second main result, we derive explicit formulas of stationary probability distribution, whose dependencies on coin parameter and the initial state of the QW are presented. A similar treatment of the limiting distribution is given by Inui et al. \cite{IKS05} for a 3-state walk governed by the Grover coin operator. A closer match to the present model is the 4-state QW with a quite different shift operator by 
Konno and Machida \cite{KM2010}, and Gettrick \cite{MG10}. As we will show in this work, the stationary probability of the QW is independent of the parity of time $t$, this is the major difference between the present model of QW and the one in \cite{KM2010}. In the case of the QW driven by many coins proposed and studied by Brun et al. \cite{BCA03}, the weak limit measures of the various scaled position operators were obtained by Segawa and Konno \cite{SK2008}. For a one-parameter family of the discrete-time QW models in both one-dimensional and two-dimensional lattices, the convergence theorems for the moments of the walker's pseudovelocity were offered in Ampadu \cite{CA11}. Some recent developments with new approaches on the asymptotic distributions and localization of discrete-time QW can be found in Cantero et al. \cite{CGMV2010}, Konno and Segawa \cite{KS11}, and Ahlbrecht et al. \cite{AVWW11}.

The rest of this paper is structured as follows: In section 2, we introduce notations and definitions for the QW that will be needed in the sequel of the paper. Section 3 contains our main results on the limiting distribution of the QW. The justifications for the results given in Section 3 are deferred to Section 4. A brief conclusion is provided in Section 5.

\section{ Formulation of quantum walks with two entangled coins on the line}

\subsection{Definitions of the quantum walks} 

For simplicity, as in \cite{V-ABBB05}, the two-coin, or more precisely, {\em two-qubit} framework is abbreviated by the name ``2cQW". Analogous to those for the single-coin (one-qubit) framework outlined in \cite{YLZ,L08}, the coin space of the 2cQW framework is the Hilbert space $\mathcal{H}_{ec}$ (``$ec$" for entangled coin) spanned by the orthonormal basis $\{|j\rangle;j\in B_c\}$ where $B_c=\{00, 01, 10, 11\}$. The position space is the Hilbert space $\mathcal{H}_p$ spanned by the orthonormal basis $\{|x\rangle;x \in \mathbb{Z} \}.$ The ``overall" state space of the system is $\mathcal{H}=  \mathcal{H}_{p} \otimes \mathcal{H}_{ec}$, in terms of which a general state of the system may be expressed by the formula:

\[\psi=\sum_{x\in\mathbb{Z}} \sum_{j\in B_c}\psi(x,j)|x\rangle\otimes|j\rangle.\]

The spatio-temporal progression of the 2cQW is governed by an ``evolution operator" $U$, which is composed of a {\em coin operator} $A_{ec}$ and a {\em shift operator} $S$.

In the 2cQW context, the coin operator is defined as the tensor product of two single-qubit operators 
$$A_{ec}=A \otimes A,$$
where $A$ may be any unitary operator acting on a single-coin space. 




The shift operator is given by:

\begin{eqnarray}
 S=|00\rangle\langle 00|\otimes \sum_i|i+1\rangle \langle i|+|01\rangle \langle 01|\otimes \sum_i|i\rangle \langle i| \nonumber \\
+ |10\rangle\langle 10|\otimes \sum_i|i\rangle \langle i|+|11\rangle \langle 11|\otimes \sum_i|i-1\rangle \langle i|. \label{equso}
\end{eqnarray}

Let $I$ denote the identity operator on $\mathcal{H}_p$. Then, in terms of $A_{ec}$ and $S$, the total evolution operator $U$ is given by 
$$U = S(I\otimes A_{ec}).$$

Given $\psi_0 \in \mathcal{H}$, where $||\psi_{0}||=1$, the expression $\psi_t= U^t \psi_0$ is called the wave function for the particle at time $t$. The corresponding 2cQW with initial state $\psi_0$ is represented by the sequence $\{ \psi_t \}_0 ^\infty$.  

Let $X$ denote the position operator on $\mathcal{H}_p$, defined by 
$X|x\rangle=x|x\rangle$ and let $\psi_t =\sum_{x \in \mathbb{Z}}\sum_{j\in B_{c}}\psi_{t}(x, j)|x\rangle\otimes |j\rangle$ be the wave function for the particle at time $t$. Then the probability $p_t(x)$ of finding the particle at the position $x$ at time $t$ is given by the standard formula 
$$p_t(x)=\sum_{j\in B_c}|\psi_t (x, j)|^2,$$
where $|\cdot|$ indicates the modulus of a complex number.
At each instant $t$, the eigenvalues of the operator $X_t\doteq {U^{\dagger}}^{t}XU^t$ equate to the possible values of the particle's position with corresponding probability $p_t(x)$.

\subsection{Fourier transform formulation of the wave function for 2cQW }

As in the single-coin setting treated in \cite{ABNVW01,GJS04}, Fourier methods can be applied in the 2cQW setting to obtain a useful formulation of the wave function. Let $\Psi_{t}^{ec}(x)\equiv [\psi_t (x, 1),\psi_t (x, 2),\psi_t (x, 3),\psi_t (x, 4)]^T$ represent the amplitude of the wave function, whose four components at position $x$ and time $t$ correspond respectively to the coin states 00, 01, 10, and 11. As usual, the superscript $T$ denotes the transpose operator. Assuming that the 2cQW is launched from the origin, then the initial quantum state of the system is reflected by the components of $\Psi_{0}^{ec}(0)=[\psi_0(0,1),\psi_0(0,2),\psi_0(0,3), \psi_0(0,4)]^T\equiv [\alpha_1, \alpha_2,\alpha_3, \alpha_4]^T$, where $\sum_{j=1}^4|\alpha_j|^2=1.$

To begin, the spatial Fourier transform of $\Psi_{t}^{ec}(x)$ is defined by
$$ \widehat{\Psi_t^{ec}}(k)=\sum_{x\in \mathbb{Z}}\Psi_{t}^{ec}(x)e^{ikx}.$$
For instance, under this transformation, the initial amplitude is related to its Fourier dual by the formula:
\begin{eqnarray}
\widehat{\Psi_{0}^{ec}}(k)=\Psi_{0}^{ec}(0).
\end{eqnarray}

In general, the Fourier dual of the overall state space of the 2cQW system is the Hilbert space $L^2(\mathbb{K})\otimes \mathcal{H}_{ec}$, consisting of $\mathbb{C}^4$-valued functions:
\begin{equation}
 \phi(k) =\left[\begin{array}{c}
  \phi_1(k)\\
   \phi_2(k) \\
    \phi_3(k) \\
     \phi_4(k)
  \end{array}\right],
\end{equation}
subject to the finiteness condition $$\Arrowvert \phi \Arrowvert^2=\Arrowvert \phi_1 \Arrowvert^2_{L^2}+\Arrowvert \phi_2 \Arrowvert^2_{L^2}+\Arrowvert \phi_3 \Arrowvert^2_{L^2}+\Arrowvert \phi_4 \Arrowvert^2_{L^2}<\infty.$$ 
Thus, given the initial state $\widehat{\Psi_{0}^{ec}}(k)$, the Fourier dual of the wave function of the 2cQW system is expressed by
\begin{equation}
\widehat{\Psi_t^{ec}}(k)=U_{ec}(k)^t\widehat{\Psi_{0}^{ec}}(k), \label{eqnwvec}
\end{equation}
where the total evolution operator $U_{ec}(k)$ on $L^2(\mathbb{K})\otimes \mathcal{H}_{ec}$ is given by 
\begin{equation}
U_{ec}(k)= \left[\begin{array}{cccc}
e^{i k} & 0 & 0 & 0\\
0       & 1 & 0 & 0\\
0       & 0 & 1 & 0\\
0       & 0 & 0 & e^{-i k}
\end{array}\right] A_{ec}\,.\label{eqnU_{ec}}
\end{equation}

Note that $U_{ec}(k)=U(k/2)\otimes U(k/2)$, where
\begin{equation}
 U(k/2)= \left[\begin{array}{cc}
e^{i k/2} & 0\\
0 & e^{-i k/2}
\end{array}\right] A\,.\label{eqnU(k)}
\end{equation}
Here \begin{equation}
 A= \left[\begin{array}{cc}
a& b\\
c& d
\end{array}\right].
\end{equation}
To wrap up this introductory section, we collect some basic facts about the eigenvalues and eigenvectors of the operators discussed above. 

Since the matrix $A$ is unitary, we may assume, without loss of generality, that its determinant is $|A|=e^{i\theta}$, where $\theta$ is a real constant. Similarly, for the unitary matrix $U(k/2)$, we may assume that its eigenvalues are $\lambda_{1}(k)=e^{i\eta(k)}$ and $\lambda_{2}(k)=e^{i(\theta-\eta(k))}$, where $\eta$ is a real-valued differentiable function of $k$. Let $v_1(k)=(v_{11}(k),v_{12}(k))^T$ and $v_2(k)=(v_{21},v_{22})^T$ denote the corresponding unit eigenvectors. Since $U_{ec}(k)=U(k/2)\otimes U(k/2)$, it follows that the eigenvalues of $U_{ec}(k)$ are:\vspace{.1in}\\
\hspace*{.3in}
\begin{equation}
\left\{
\begin{array}{l}
\Lambda_1(k)=[\lambda_1(k)]^2=e^{i\varphi(k)}$ where $\varphi(k)=2\eta(k)\\
\Lambda_2(k)=\lambda_1 (k)\lambda_2(k)=|A|=e^{i\theta}\\
\Lambda_3(k)=\lambda_1 (k)\lambda_2(k)=|A|=e^{i\theta}\\
\Lambda_4(k)=[\lambda_2(k)]^2=e^{i(2\theta-\varphi(k))}.
\end{array}\right. \label{eigen-1}
\end{equation}

Correspondingly, the unit eigenvectors are:\vspace{.1in}\\
\hspace*{.3in}$\left\{
\begin{array}{l}
V_1(k)=v_1(k)\otimes v_{1}(k)\\ 
V_2(k)=v_1(k)\otimes v_{2}(k)\\ 
V_3(k)=v_2(k)\otimes v_{1}(k)\\ 
V_4(k)=v_2(k)\otimes v_{2}(k). 
\end{array}\right.$

Finally, in terms of eigenvalues and eigenvectors, the wave function $ \widehat{\Psi_t^{ec}}(k)$ may be expanded as follows:
\begin{eqnarray}
\widehat{\Psi_t^{ec}}(k)&=& U_{ec}^t(k)\widehat{\Psi_0^{ec}}(k)\nonumber\\
\mbox{}&=& e^{it\varphi(k)}\langle V_1(k),\widehat{\Psi_0^{ec}}(k)\rangle V_1(k)\nonumber\\
\mbox{}&\mbox{}&+e^{it\theta}\langle V_2(k),\widehat{\Psi_0^{ec}}(k)\rangle V_2(k)\nonumber\\
\mbox{}&\mbox{}&+e^{it\theta}\langle V_3(k),\widehat{\Psi_0^{ec}}(k)\rangle V_3(k)\nonumber\\
\mbox{}&\mbox{}&+e^{it(2\theta-\varphi(k))}\langle V_4(k),\widehat{\Psi_0^{ec}}(k) \rangle V_4(k).\label{eqnwaec}
\end{eqnarray}


\section{Limiting distributions for 2cQW}


To prepare for deriving the limiting distribution of 2cQW, we present some properties of the evolution operator $U_{ec}(k)$.
As $\overline{\lambda} D\lambda$ does in the case of the single-coin QW \cite{L08}, the expression $\overline{\Lambda}D\Lambda$ acts as an essential ingredient in the formulae for the moments of position probability distribution of 2cQW. What follows is the relationship between $\overline{\Lambda}D\Lambda$ and $\overline{\lambda} D\lambda$, which is easily verified based on  Eq. (\ref{eigen-1}).

\begin{equation}
\left\{
\begin{array}{l}
\overline{\Lambda_1}D\Lambda_1(k)=2\overline{\lambda_1}D\lambda_1(k)\\
\overline{\Lambda_2}D\Lambda_2(k)=\overline{\Lambda_3}D\lambda_3=0\\
\overline{\Lambda_4}D\Lambda_4(k)=2\overline{\lambda_2}D\lambda_2(k).
\end{array}\right. \label{keyingredient}
\end{equation}

According to \cite{L08}, we have 

\begin{equation}
\left\{
\begin{array}{l}
v_{i1}\overline{v_{i1}}=\frac{1+2\overline{\lambda_i}D\lambda_i}{2}\\
v_{i2}\overline{v_{i2}}=\frac{1-2\overline{\lambda_i}D\lambda_i}{2}\\
\overline{\lambda_1}D\lambda_1+\overline{\lambda_2}D\lambda_2=0.
\end{array}\right. \label{vec.-eigen.}
\end{equation}

 Eq. (\ref{vec.-eigen.}) implies that $v_{21}\overline{v_{21}}=v_{12}\overline{v_{12}}$ and $v_{22}\overline{v_{22}}=v_{11}\overline{v_{11}}$.
 
In addition, we can show the validity of the following identity by a simple linear algebra.
\begin{eqnarray}
v_{i1}\overline{v_{i2}}=\frac{\overline{c}\lambda_iv_{i1}\overline{v_{i1}}-b\overline{\lambda_i}v_{i2}\overline{v_{i2}}}{a\overline{\lambda_i}-\overline{d}\lambda_i}. \label{vec-matrix}
\end{eqnarray}

For better presentation of our analysis of the limiting distribution of the 2cQW, we specify the coin operator matrix $A$ as follows,
 \begin{equation}
 A(\beta)= \left[\begin{array}{cc}
\cos \beta& \sin \beta \\
\sin \beta& -\cos \beta
\end{array}\right].\label{matrixa}
\end{equation}
Here $\beta \in (0,\frac{\pi}{2})$. The corresponding evolution operator is restated here for convenience.

\begin{equation}
U(k/2)=
\left[ \begin{array}{ccc}
e^{i k/2}\cos \beta  & e^{i k/2}\sin \beta \\
e^{-i k/2}\sin \beta &-e^{-i k/2}\cos \beta \\
\end{array}\right].
\end{equation}

\quad By straightforward calculation, the eigenvalues are 
\begin{equation}
\lambda_{j}(k)= \pm \sqrt{1-\cos^2\beta \sin^2(k/2)}+i \cos \beta \sin (k/2),\label{eigenvalue}
\end{equation}
so that 

\begin{equation}
\overline{\lambda_{1}}D\lambda_{1}=(\cos \beta \cos (k/2))/(2\sqrt{1-\cos^2 \beta \sin^2 (k/2)}).\label{keyingredient2}
\end{equation}

By Eqs. {(\ref{keyingredient}), {(\ref{vec.-eigen.})} and (\ref{keyingredient2}), one can obtain
\begin{eqnarray}
\overline{\Lambda_1}D\Lambda_1(k)=-\overline{\Lambda_4}D\Lambda_4(k)=\frac{\cos \beta\cos \frac{k}{2}}{\sqrt{\sin^2\beta+\cos^2\beta\cos^2\frac{k}{2}}}. \label{eigen-beta}
\end{eqnarray}

 Eqs. (\ref{vec.-eigen.}), (\ref{vec-matrix}) and (\ref{eigenvalue}) together imply the identities:
\begin{eqnarray}
v_{i1}\overline{v_{i2}}=\tan \beta[\frac{\lambda_i-\overline{\lambda_i}}{2(\lambda_i+\overline{\lambda_i})}+\overline{\lambda_i}D\lambda_i],\, v_{21}\overline{v_{22}}=-v_{11}\overline{v_{12}}. \label{vec-matrix-b}
\end{eqnarray}

According to the methods by Grimmett {\it et al.} \cite{GJS04}, the moments of the position distribution are given as 
\begin{eqnarray}
E(X_t^r)=\int_{0}^{2\pi}\langle \widehat{\Psi_t^{ec}}(k), D^r \widehat{\Psi_t^{ec}}(k) \rangle \frac{d k}{2\pi}
\end{eqnarray}
where $D=-\mathrm {i} \mathrm{d}/\mathrm{d} k$ denote the position operator in $k$-space.

Using the standard calculations, we arrive at, as $t\rightarrow \infty$,
\begin{eqnarray}
E[(X_t/t)^r]=\int_0^{2\pi}\sum_j(\frac{D\Lambda_j(k)}{\Lambda_j(k)})^r|\langle V_j(k), \widehat{\Psi_0^{ec}}(k)\rangle|^2\frac{dk}{2\pi}+O(t^{-1}).
\end{eqnarray}

By the method of moments (see \cite{GJS04} and references therein), we may derive the following limit theorem.

Theorem 1. \, Suppose the 2cQW, controlled by the coin operator $A^{\otimes 2}$, is launched from the origin in the initial state $\Psi_{0}^{ec}(0)=\alpha_1|00\rangle+\alpha_2|01\rangle+\alpha_3|10\rangle+\alpha_4|11\rangle$, where $\sum_{j=1}^4|\alpha_j|^2=1$. For $y\in [-1,1]$, let $\delta_0(y)$ denote the \textit{point mass at the origin} and let $I_{(a,b)}(y)$ denote the \textit{indicator function} of the real interval $(a,b)$. Then, as $t\rightarrow \infty$, the normalized position distribution $f_{t}(y)$ associated with $\frac{1}{t}X_{t}$ converges, in the sense of a weak limit, to the density function
\vskip -0.5cm
\begin{eqnarray}
f(y)=c_{00}\delta_0(y)+\frac{\tan\beta I_{(-\cos\beta, \cos\beta)}(y)}{\pi (1-y^2)\sqrt{1-y^2\sec^2\beta}}\sum_{j=0}^2c_j y^j. \label{densityfunc}
\end{eqnarray}
\vskip -0.2cm
In the above formula, the coefficients $c_{00}$, $c_{0}$, $c_{1}$ and $c_{2}$ are given by  
\vskip -0.5cm
\begin{equation}
\left\{
\begin{array}{l}
c_{00}=\frac{\sin \beta}{2}-(\sin \beta-1)(|\alpha_2|^2+|\alpha_3|^2)+\tan^2\beta(\frac{1}{\sqrt{\sin\beta}}-\sqrt{\sin\beta})^2\mathrm{Re}(\alpha_1\overline{\alpha_4}) \nonumber\\
\quad\quad\quad+(\sin\beta-1)\tan\beta\mathrm{Re}(\alpha_1\overline{\alpha_2}+\alpha_1\overline{\alpha_3}-\alpha_2\overline{\alpha_4}-\alpha_3\overline{\alpha_4})-\sin\beta\mathrm{Re}(\alpha_2\overline{\alpha_3}), \nonumber\\
c_2=\frac{1}{2}-(|\alpha_2|^2+|\alpha_3|^2)-\mathrm{Re}(\alpha_2\overline{\alpha_3})+(2\tan^2\beta+1)\mathrm{Re}(\alpha_1\overline{\alpha_4}) \nonumber\\
\quad\quad+\tan\beta\mathrm{Re}(\alpha_1\overline{\alpha_2}+\alpha_1\overline{\alpha_3}-\alpha_2\overline{\alpha_4}-\alpha_3\overline{\alpha_4}),\nonumber\\
c_1=|\alpha_1|^2-|\alpha_4|^2+\tan\beta\mathrm{Re}(\alpha_1\overline{\alpha_2}+\alpha_1\overline{\alpha_3}+\alpha_2\overline{\alpha_4}+\alpha_3\overline{\alpha_4}),\nonumber\\
c_0=\frac{1}{2}+\mathrm{Re}(\alpha_2\overline{\alpha_3}-\alpha_1\overline{\alpha_4}).\nonumber
\end{array}\right.
\end{equation}
Where $\mathrm{Re}(z)$ is the real part of a complex number $z$.
\vskip0.1in

As an example, we consider the case when  $\beta=\frac{\pi}{4}$, $\alpha_1=\alpha_4=\frac{\sqrt{2}}{2}$, and $\alpha_2=\alpha_3=0$. Direct calculations shows that the density function in Theorem 1 becomes 
 \begin{eqnarray}
f(y)=(\sqrt{2}-1)\delta_0(y)+2y^2\frac{I_{(-\frac{\sqrt{2}}{2}, \frac{\sqrt{2}}{2})}(y)}{\pi (1-y^2)\sqrt{1-2y^2}}. \label{densityfunc-spe}
\end{eqnarray}

The component with delta function in Eq. (\ref{densityfunc}) is an indicator of the phenomonon of ``localization". In contrast, the density function for weak limit measure of the basic model of QW, i.e., 2-state QW, does not comprise the delta function. In fact, in the pioneered papers for the treatment of weak limit measure of QW, Konno \cite{K02,K2005} and Grimmett {\it et al.} \cite{GJS04} obtained a limiting density function of the normalized position operator $\frac{X_t}{t}$ associated with 2-state Hadamard QW as follows:
\begin{eqnarray}
f(y)=\frac{I_{(-\frac{\sqrt{2}}{2}, \frac{\sqrt{2}}{2})}(y)}{\pi (1-y^2)\sqrt{1-2y^2}}. \label{densityfunc-spe}
\end{eqnarray}

It should be noted that $c_{00}\ne 0$ is the {\it necessary and sufficient} condition for localization of the QW on both the quantum coin and the initial state. Localization does not always occur in 2cQW. For instance, when $\beta=\frac{\pi}{4}$, $\alpha_1=\alpha_2=\alpha_3=-\frac{1}{2}$ and $\alpha_4=\frac{1}{2}$, then $c_{00}=0$ by Theorem 1, and the density function of Eq.(\ref{densityfunc}) just becomes the one given in Eq.(\ref{densityfunc-spe}). Interestingly, the pattern of function in Eq.(\ref{densityfunc-spe}) and/or delta function turn out to be generic components of the density functions for many existing models of QW. Examples of these include, the three-state Grover walks by Inui et al. \cite{IKS05}, the QW with multiple coins by Segawa and Konno \cite{SK2008}, and the four-state QW with a four-direction shift operator by Konno and Machida \cite{KM2010}(we point out that this is a closer model to 2cQW considered in the present paper, which actually is a four-state QW with a three-direction shift operator.)

Finally, it may be worth noticing that the coefficient $c_{00}$ in Eq. (\ref{densityfunc}), is the sum of the stationary probability $\lim_{t\rightarrow \infty}p_t(x)$, i.e., $c_{00}=\sum_{x\in \mathbb{Z}}\lim _{t\rightarrow \infty}p_t(x)$ (see Theorem 2 in \cite{LP09}). Our analysis of the asymptotic distributions of 2cQW would not be complete without a consideration of the stationary probability of the QW. By using complex integrals as did by Konno and Machida \cite{KM2010}, we obtain the following explicit formulas for the stationary probability.

Theorem 2. \, Suppose the 2cQW, controlled by the coin operator $A^{\otimes 2}$, is launched from the origin in the initial state $\Psi_{0}^{ec}(0)=\alpha_1|00\rangle+\alpha_2|01\rangle+\alpha_3|10\rangle+\alpha_4|11\rangle$, where $\sum_{j=1}^4|\alpha_j|^2=1$.  Let $p(x)=\lim_{t\rightarrow \infty}p_t(x)$. Then the limiting probability of finding the walker at $|x\rangle$ is given below.
\vskip -0.5cm
\begin{eqnarray} \!\!\!\mathrm{(i)}\,p(0)
\!\!&=&\!\!\tan^2\beta\sec^2\beta(1-\sin\beta)^2(|\alpha_1|^2+|\alpha_4|^2)+\sec^2\beta(1-\sin\beta)(|\alpha_2|^2+|\alpha_3|^2)\nonumber\\
\!\!&+&\!\!\tan\beta\sec^2\beta(1-\sin\beta)^2\mathrm{Re}(\alpha_2\overline{\alpha}_4+\alpha_3\overline{\alpha}_4-\alpha_1\overline{\alpha}_2-\alpha_1\overline{\alpha}_3)\nonumber\\
\!\!&-&\!\!2\tan\beta\sec\beta(1-\sin\beta)\mathrm{Re}(\alpha_2\overline{\alpha}_3).\label{pro-at-0}
\end{eqnarray}
\vskip -0.2cm
\hskip -0.4cm (ii)\,$p(x)=J_{+}(\alpha,\beta)[(1-\sin\beta)^4\sec^4\beta]^x$ for $x\ge 1$ and 

\hskip -0.5cm (iii)\,$p(x)=J_{-}(\alpha,\beta)[(1-\sin\beta)^4\sec^4\beta]^{-x}$ for $x\le -1.$ 
\vskip -0.6cm
\begin{eqnarray}\mathrm{Here}\,\,\,\,
\!\!\!J_{+}(\alpha,\beta)
\!\!&=&\!\!\tan^2\beta[ \sec^2\beta (1+\sin\beta)^2|\alpha_1|^2+ |\alpha_2|^2+|\alpha_3|^2+\sec^2\beta(1-\sin\beta)^2|\alpha_4|^2\nonumber\\
\!\!&-&\!\!2\sec\beta(1+\sin\beta)\mathrm{Re}(\alpha_1\overline{\alpha_2}+\alpha_1\overline{\alpha_3})+2\mathrm{Re}(\alpha_1\overline{\alpha_4}+\alpha_2\overline{\alpha_3})\nonumber\\
\!\!&-&\!\!2\sec\beta(1-\sin\beta)\mathrm{Re}(\alpha_2\overline{\alpha_4}+\alpha_3\overline{\alpha_4})].\\
\!\!\!J_{-}(\alpha,\beta)
\!\!&=&\!\!\tan^2\beta[ \sec^2\beta (1-\sin\beta)^2|\alpha_1|^2+ |\alpha_2|^2+|\alpha_3|^2+\sec^2\beta(1+\sin\beta)^2|\alpha_4|^2\nonumber\\
\!\!&+&\!\!2\sec\beta(1-\sin\beta)\mathrm{Re}(\alpha_1\overline{\alpha_2}+\alpha_1\overline{\alpha_3})+2\mathrm{Re}(\alpha_1\overline{\alpha_4}+\alpha_2\overline{\alpha_3})\nonumber\\
\!\!&+&\!\!2\sec\beta(1+\sin\beta)\mathrm{Re}(\alpha_2\overline{\alpha_4}+\alpha_3\overline{\alpha_4})].
\end{eqnarray}

Let's revisit the first example where $\beta=\frac{\pi}{4}$, $\alpha_1=\alpha_4=\frac{\sqrt{2}}{2}$ and $\alpha_2=\alpha_3=0$, we have $p(0)=3-2\sqrt{2}$ by Eq. (\ref{pro-at-0}). This agrees with the findings reported in \cite{V-ABBB05,LP09}. Moreover, direct calculations by Theorem 2 show that $\sum_{x\in \mathbb{Z}}p(x)=\sqrt{2}-1$, which is consistent with the previous result by Theorem 1. In this case, localization occurs. In the other case when $\beta=\frac{\pi}{4}$, $\alpha_1=\alpha_2=\alpha_3=-\frac{1}{2}$ and $\alpha_4=\frac{1}{2}$, it can be deduced that $p(x)=0$ for any integer $x$ from Theorem 2, thus $c_{00}=\sum_{x\in \mathbb{Z}}p(x)=0$ which is also consistent with our previous calculation from Theorem 1. Therefore, localization, in this scenario, does not take place. Similar conclusions hold for the 4-state QW given in \cite{KM2010}. However, unlike that QW in \cite{KM2010}, the stationary distribution for 2cQW is independent of the parity of time $t$, this is one important difference that distinguishes one from the other.



\section{Proof of theorems}
This section is devoted to the justification for the limit theorems presented in the preceding section. 

Proof of Theorem 1.\, We begin with the moments of the position distribution:
\begin{eqnarray}
E[(X_t/t)^r]=\int_0^{2\pi}\sum_j(\frac{D\Lambda_j(k)}{\Lambda_j(k)})^r|\langle V_j(k), \widehat{\Psi_0^{ec}}(k)\rangle|^2\frac{dk}{2\pi}+O(t^{-1}).
\end{eqnarray}

By the method of moments (see \cite{GJS04} and references therein), the weak limit of $X_t/t$ exists. Let $Y$ be this weak limit. Then we have 
\begin{eqnarray}
\mathrm{P}(Y \le y)=\int_{h^{-1}(k,j)((-\infty, y])}\sum_{j=1}^4|\langle V_j(k), \widehat{\Psi_0^{ec}}(k)\rangle|^2\frac{dk}{2\pi}\label{density-fun}
\end{eqnarray}
where $h(k,j)=\overline{\Lambda_j}D\Lambda_j(k).$ 

According to  Eqs. {(\ref{keyingredient}) and {(\ref{eigen-beta})}, the probability distribution function in Eq. (\ref{density-fun}) can be written as 
\begin{eqnarray}
\!\!\!\mathrm{P}(Y \le y)
\!\!&=&\!\!\int_{2\arccos(\tan\beta \frac{y}{\sqrt{1-y^2}})}^{2\pi}|\langle V_1(k), \widehat{\Psi_0^{ec}}(k)\rangle|^2\frac{dk}{2\pi}\nonumber\\
\!\!&+&\!\!H(y)\int_0^{2\pi}\{|\langle V_2(k), \widehat{\Psi_0^{ec}}(k)\rangle|^2+|\langle V_3(k), \widehat{\Psi_0^{ec}}(k)\rangle|^2\}\frac{dk}{2\pi}\nonumber\\
\!\!&+&\!\!\int_0^{2\arccos(-\tan \beta \frac{y}{\sqrt{1-y^2}})}|\langle V_4(k), \widehat{\Psi_0^{ec}}(k)\rangle|^2\frac{dk}{2\pi}. \label{distribution}
\end{eqnarray}
Here $H(y)$ is {\em Heaviside function}, which is the cumulative distribution function of $\delta_0(y)$.

After taking derivatives of both sides of Eq. (\ref{distribution}) with respect to $y$, we obtain the density as follows
\begin{eqnarray}
\!\!\!f(y)
\!\!&=&\!\!\frac{d\mathrm{P}(Y\le y)}{dy}\nonumber\\
\!\!&=&\!\!\frac{\tan\beta}{\pi(1-y^2)\sqrt{1-y^2\sec^2\beta}}|\langle V_1(k), \widehat{\Psi_0^{ec}}(k)\rangle|^2_{k=2\arccos(\tan\beta \frac{y}{\sqrt{1-y^2}})} \label{density1} \nonumber\\ 
\!\!&+&\!\!\delta_0(y)\int_0^{2\pi}\{|\langle V_2(k), \widehat{\Psi_0^{ec}}(k)\rangle|^2+|\langle V_3(k), \widehat{\Psi_0^{ec}}(k)\rangle|^2\}\frac{dk}{2\pi}  \label{density2} \nonumber\\
\!\!&+&\!\!\frac{\tan\beta}{\pi(1-y^2)\sqrt{1-y^2\sec^2\beta}}|\langle V_4(k), \widehat{\Psi_0^{ec}}(k)\rangle|^2_{k=2\arccos(-\tan\beta \frac{y}{\sqrt{1-y^2}})}   \label{density}
\end{eqnarray}

Applying Eqs. (\ref{vec.-eigen.}), (\ref{eigenvalue}) ,  (\ref{keyingredient2}) and (\ref{vec-matrix-b}) to simplify Eq. (\ref{density}), one can obtain the density function $f(y)$ given in Eq. (\ref{densityfunc})

\vskip0.1in
 
Proof of Theorem 2.\, By Eqs. (\ref{eigen-1}) and (\ref{matrixa}), the eigenvalues of $U_{ec}(k)$ can be obtained as 

$\Lambda_1=\lambda_1^2$, $\Lambda_2=\Lambda_3=-1$, $\Lambda_4=\lambda_2^2$ where $\lambda_1$ and $\lambda_2$ are given by Eq. (\ref{eigenvalue}).

By inverse Fourier transformation, the amplitude of the wave function of the walker at the position $x$ and the time $t$ is given by 
\begin{eqnarray}
\Psi_t^{ec}(x) &=& [\psi_t(x,1),\psi_t(x,2),\psi_t(x,3), \psi_t(x,4)]^T \nonumber \\
\mbox{} &=& \int_0^{2\pi}e^{-ixk}\widehat{\Psi_t^{ec}}(k)\frac{dk}{2\pi}\nonumber\\
\mbox{} &=& \sum_{j=1}^4\int_0^{2\pi}e^{-ixk}\Lambda_j^t\langle V_j(k),\widehat{\Psi_0^{ec}}(k)\rangle V_j(k), \quad 
\end{eqnarray}

Using the method of stationary phase as used in \cite{LP09}, we may deduce that 
\begin{eqnarray}
\Psi_t^{ec}(x) \thicksim  \int_0^{2\pi}e^{-ixk}(-1)^t\sum_{j=2}^3\langle V_j(k),\widehat{\Psi_0^{ec}}(k)\rangle V_j(k)\frac{dk}{2\pi}. \label{complexintegral}
\end{eqnarray}

The right hand side of Eq. (\ref{complexintegral}) can be converted into complex integrals,  and the exact values for the complex integrals can be computed. Therefore we reach
\begin{equation}
 \Psi_t^{ec}(0)\thicksim(-1)^t\left[\begin{array}{c}
 \frac{c_1z_{1}+b_1}{z_{1}-z_{2}} \\
  \alpha_2-\alpha_1\tan \beta +\frac{c_2z_{1}+b_2}{z_{1}-z_{2}} \\
    \alpha_3-\alpha_1\tan \beta+\frac{c_3z_{1}+b_3}{z_{1}-z_{2}}\\
     (\alpha_2+\alpha_3)\tan\beta-2\alpha_1\tan^2\beta+\frac{c_4z_{1}+b_4}{z_{1}-z_{2}}
  \end{array}\right].\label{p(0)}
\end{equation}

For $x\ge 1$,
\begin{eqnarray}
\Psi_t^{ec}(x)\thicksim(-1)^t\frac{1}{z_2^{x+1}}[\frac{b_1z_2+c_1}{z_1-z_2},\frac{b_2z_2+c_2}{z_1-z_2},\frac{b_3z_2+c_3}{z_1-z_2},\frac{b_4z_2+c_4}{z_1-z_2}].\label{p(x)+}
\end{eqnarray}

For $x\le -1$,
\begin{eqnarray}
\Psi_t^{ec}(x)\thicksim(-1)^t\frac{1}{z_1^{x+1}}[\frac{b_1z_1+c_1}{z_1-z_2},\frac{b_2z_1+c_2}{z_1-z_2},\frac{b_3z_1+c_3}{z_1-z_2},\frac{b_4z_1+c_4}{z_1-z_2}].\label{p(x)-}
\end{eqnarray}

Here $z_1=-(1-\sin \beta)^2\sec^2\beta$, $z_2=-(1+\sin \beta)^2\sec^2\beta$; and 

\begin{equation}
\left\{
\begin{array}{l}
b_1=-(\alpha_2+\alpha_3)\tan \beta+2\alpha_1\tan^2 \beta\nonumber\\
c_1=-(\alpha_2+\alpha_3)\tan \beta-2\alpha_4\tan^2 \beta \nonumber\\
b_2=(\alpha_1+\alpha_4)\tan \beta-2(\alpha_2+\alpha_3)\tan^2\beta+4\alpha_1\tan^3\beta \nonumber\\
c_2=(\alpha_1+\alpha_4)\tan \beta \nonumber\\
b_3=(\alpha_1+\alpha_4)\tan \beta-2(\alpha_2+\alpha_3)\tan^2\beta+4\alpha_1\tan^3\beta \nonumber\\
c_3=(\alpha_1+\alpha_4)\tan \beta \nonumber\\
b_4=-(\alpha_2+\alpha_3)\tan\beta+(4\alpha_1+2\alpha_4)\tan^2\beta-4(\alpha_2+\alpha_3)\tan^3\beta+8\alpha_1\tan^4\beta \nonumber\\
c_4=-(\alpha_2+\alpha_3)\tan \beta+2\alpha_1\tan^2 \beta
\end{array}\right.
\end{equation}

The formulas for the stationary probability can be derived from  Eqs. (\ref{p(0)}), (\ref{p(x)+}) and (\ref{p(x)-}).

\section{Conclusions}

For the quantum walk driven by two coins introduced by \cite{V-ABBB05}, we here offer thoroughly analytic treatments of the asymptotic behaviors of position probability distributions. The fact that the walker stalls at each time step with a fixed positive probability as indicated by the three-direction shift operator, does not imply that the stationary probability must be nonzero and localization will surely occur. As shown by the example following Theorem 1, localization may not occur for some initial state and some choice of the coin parameter; therefore, the density function of the weak limit of $\frac{X_t}{t}$ in Theorem 1, in this case, does not contain the delta function as one of its components. On the other hand, in the case where localization occurs, we show in Theorem 2 that the stationary probability exponentially decays with the absolute value of the walker's position, independent of the parity of time. Finally, we should point out that Theorem 2 in the present paper refines on a previous result (see Theorem 1 in \cite{LP09}), which also concerns the stationary distribution. Indeed, Theorem 2 is a stronger version of Theorem 1 in \cite{LP09}. We wish that Theorem 2 would clarify and correct the comments in \cite{LP09} on the decay rate of the stationary distribution for 2cQW.

\section*{Acknowledgment}

The research was supported by NSF grant CCF-1005564.

\end{document}